\documentclass[apj,twocolumn]{emulateapj}
\usepackage{amsmath}
\usepackage{txfonts}
\usepackage{amssymb}
\usepackage{graphicx}
\usepackage{textcomp}
\usepackage{color}
\usepackage{hyperref}
\renewcommand\email\texttt
\newcommand\beq{\begin{equation}}
\newcommand\eeq{\end{equation}}

\begin{document} 

\shorttitle{An extended view of the Pisces overdensity}
\shortauthors{J. D.  Nie et al.}
\title{An extended view of the Pisces overdensity from the SCUSS Survey}
\author{
J. D. Nie\altaffilmark{1},
M. C. Smith\altaffilmark{2},
V. Belokurov\altaffilmark{3},
X.H. Fan\altaffilmark{4},
Z. Fan\altaffilmark{1},
M.J. Irwin\altaffilmark{3},
Z.J. Jiang\altaffilmark{1},
Y.P. Jing\altaffilmark{5}
S.E. Koposov\altaffilmark{3},
M. Lesser\altaffilmark{4}
J. Ma\altaffilmark{1},
S.Y. Shen\altaffilmark{2},
J.L. Wang\altaffilmark{1},
Z.Y. Wu\altaffilmark{1},
T.M. Zhang\altaffilmark{1},
X. Zhou\altaffilmark{1},
Z.M. Zhou\altaffilmark{1},
H. Zou\altaffilmark{1}
}
\altaffiltext{1}{Key Laboratory of Optical Astronomy, National Astronomical Observatories, 
Chinese Academy of Sciences, Beijing 100012, China;\email{jdnie@bao.ac.cn}}
\altaffiltext{2}{Shanghai Astronomical Observatory, 80 Nandan Road, Shanghai 200030, China;
\email{dr.mcsmith@me.com}}
\altaffiltext{3}{Institute of Astronomy, Madingley Road, Cambridge, CB3 0HA, UK}
\altaffiltext{4}{Steward Observatory, University of Arizona, Tucson, AZ 85721}
\altaffiltext{5}{Center for Astronomy and Astrophysics, Department of Physics and Astronomy, Shanghai Jiao Tong University, Shanghai 200240, China}

\begin{abstract}
SCUSS is a $u$-band photometric survey covering about 4000 square
degree of the South Galactic Cap, reaching depths of up to 23 mag. 
By extending around 1.5 mag deeper than SDSS single-epoch $u$ data,
SCUSS is able to probe much a larger volume of the outer halo,
i.e. with SCUSS data blue horizontal branch (BHB) stars can trace the 
outer halo of the Milky Way as far as 100--150 kpc. Utilizing this
advantage we combine SCUSS $u$ band with SDSS DR9 $g r i$ photometric
bands to identify BHB stars and explore halo substructures.
We confirm the existence of the Pisces overdensity, which is a
structure in the outer halo (at around 80 kpc) that was discovered
using RR Lyrae stars. For the first time we are able to determine its
spatial extent, finding that it appears to be part of a stream with a
clear distance gradient. The stream, which is $\sim$5 degrees wide
and stretches along $\sim$25 degrees, consists of 20--30
BHBs with a total significance of around 6$\sigma$ over the background.
Assuming we have detected the entire stream and that the
progenitor has fully disrupted, then the number of BHBs suggests the
original system was similar to smaller classical or a larger ultra-faint dwarf galaxy.
On the other hand, if the progenitor still exists, it can be
hunted for by reconstructing its orbit from the distance gradient of
the stream. This new picture of the Pisces overdensity sheds new light
on the origin of this intriguing system.
\end{abstract}


\section{Introduction}

The Pisces overdensity is a recent discovery, being one of the most distant 
substructures in the Galactic halo. This substructure was first 
uncovered by \citet{2007AJ....134.2236S} using RR Lyraes from
the multi-epoch Sloan Digital Sky Survey (SDSS) Stripe 82
data, along with a number of other candidate overdensities. Located at
RA $\sim$354$^\circ$ and Dec $\sim$0$^\circ$, with a median heliocentric
distance of 81 kpc, this was named `Structure J'.   
Subsequently, \citet{2009MNRAS.398.1757W} used the same data, this
time analyzing the light-curves from \citet{Bramich2008},
and independently found what appears  to be the same structure, 
located at $-20^{\circ}<$RA$<0^{\circ}$, $-1.25^{\circ}<$Dec$<1.25^{\circ}$ 
with a heliocentric distance of 80 kpc. They named it the `Pisces
overdensity' after the constellation in which it is located. To
confirm whether this photometric overdensity is truly a coherent
structure, as opposed to a chance concentration,
\citet{2009ApJ...705L.158K} obtained spectroscopy for eight RR Lyrae
stars in the Pisces overdensity region and found five of them have a
narrow range of velocities, which suggests that the overdensity is
genuine. Later \citet{2010ApJ...717..133S} observed a further four RR
Lyrae stars and, using the combined sample of 12 stars, confirmed the
presence of a secondary velocity structure, which was tentatively
found in the original \citet{2009ApJ...705L.158K} sample.

In terms of its extension, one of the most intriguing aspects of the
Pisces overdensity is that it appears to be distributed over a large
area on the sky. The original detection spans 10--15$^\circ$, which
at a distance of 80 kpc corresponds to a width of 15--20 kpc. However,
since the existing studies are based only on the thin 2.5$^\circ$-wide
stripe Stripe 82 data, it is impossible to determine the full
extent. This makes it hard to draw any firm conclusions as to
the nature of the progenitor, beyond the fact that it appears more
likely to be a tidally disrupted galaxy rather than an intact system.
The full extent can only be determined by either deep wide-field
photometric surveys or spectroscopy of faint halo tracers at distances
consistent with the Pisces overdensity.

In this study, we will use a deep photometric sky survey, called the
South Galactic Cap $u$ band Sky Survey (SCUSS), to investigate the
extension of the Pisces overdensity. SCUSS is a $u$-band (wavelength
$\sim$3538$\AA$) photometric survey, covering around 4000 square
degree (sq. deg.) in the South Galactic Cap region. 80\% of the area
overlaps with the southern SDSS data, but the SCUSS magnitude limit is
1--1.5 mag deeper than SDSS III DR9 $u$-band data (wavelength
$\sim$3551$\AA$). This deep SCUSS $u$-band data can probe a much
larger volume of the halo, easily reaching beyond 100 kpc for Blue
Horizontal Branch (BHB) stars.
Since the Pisces overdensity is about 80 kpc away, the depth and sky
coverage of SCUSS is ideal to study the spatial extent of this
structure.

We organize the paper as follows: Section \ref{data} introduces the
SCUSS data; in Section \ref{selection} we select BHB stars from the
SCUSS survey; in Section \ref{pisces} we investigate the spatial extent
of the Pisces overdensity using these BHBs; and in Section
\ref{giant_branch} we explore the distribution of giant branch stars;
finally we present a summary and discussion in Section \ref{conclusion}.

\section{Data overview}\label{data}
We mainly utilize the SCUSS $u$-band\footnote{
In the following, unless otherwise stated, when we discuss the
$u$-band magnitude this corresponds to the SCUSS magnitude.} and SDSS DR9
single-epoch $gri$ bands data in this study. SCUSS is an international
collaboration between the National Astronomical Observatories of China
(Chinese Academy of Sciences) and Steward Observatory (University of
Arizona, USA). This survey has imaged $\sim$4000 sq. deg. of the
Southern Galactic Cap, with galactic latitude  $b<-30^{\circ}$ and
equatorial latitude Dec$>-10^{\circ}$ at a wavelength 
of $u\sim$3538$\AA$. The survey was carried out using the 90Prime imager of 
the 2.3m Bok telescope at Steward Observatory on Kitt Peak between 2010 and 
2013. The detector consists of an array of four 4K$\times$4K (64-megapixel) 
CCDs and the field of view is $1.08^{\circ}\times1.03^{\circ}$, resulting 
in a pixel resolution of 0.454$\arcsec$. The typical seeing during the
observing period is about 2.0$\arcsec$ and the exposure time of each
image is 300s. To obtain the required magnitude limit two dithered
exposures for each field were taken and if any fields did not meet the
quality requirement (for example due to bad weather), extra exposures
were added. More details about the SCUSS survey can be found in
\citet{Zhou2015} and \citet{Zou2015}.

\begin{figure}
\center
\includegraphics[scale=.4]{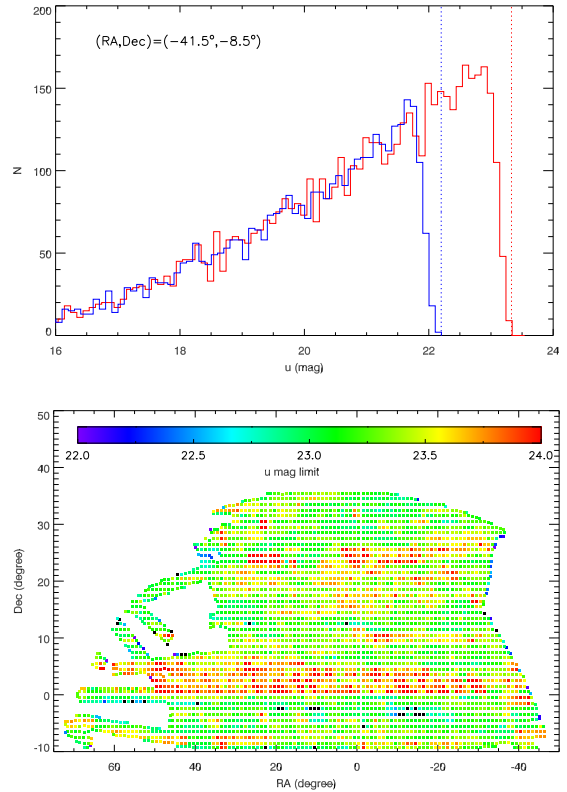}
\center
\caption{An estimation of the SCUSS magnitude limit. The top panel
shows an example of how we define the magnitude limit. The red
histogram shows the luminosity function in one SCUSS field (located
at RA = -41.5$^\circ$ and Dec = -8.5$^\circ$), while the blue histogram is
for SDSS data in the same field. We use PSF magnitudes and only
consider point sources with $u$-band error less than 0.2 mag (5$\sigma$).
We take the maximum magnitude (vertical dashed lines), as our
definition of the magnitude limit. The bottom panel shows the
resulting SCUSS magnitude limit for each field in the 3400
sq. deg. overlap region of SCUSS and SDSS.}
\center
\label{maglimit}
\end{figure}

The SCUSS observing strategy yields a significant increase in
depth compared to single-epoch SDSS data. We compare these two surveys
in Figure \ref{maglimit}, where SCUSS co-added PSF magnitudes are
compared with SDSS PSF magnitudes. In this figure we divide the whole
SCUSS footprint into $1.08^{\circ}\times1.03^{\circ}$ bins and analyze
the magnitude distribution (i.e. luminosity function) of point sources
in each bin, where star-galaxy separation is based on the SDSS
classification. The luminosity function turns over at faint magnitudes
as the survey reaches its detection limit and becomes incomplete.
In our case the limiting magnitude is defined as the maximum magnitude
of a star with a given error. For both SDSS \& SCUSS datasets we
consider detections with $u$-band photometric error less than 0.2 mag
(5$\sigma$). This procedure is illustrated in the top panel of Figure
\ref{maglimit}. For this example field (located at
$\rm RA = -41.5^\circ, Dec = -8.5^\circ$) we find that the SCUSS
magnitude limit is around 23.3, which is around 1 mag deeper than
SDSS. The bottom panel of this figure shows the distribution of
magnitude limit across the SCUSS footprint. In general the SCUSS
fields are 1 to 1.5 mag deeper. As our BHB classification
requires $ugri$ photometry, we have only analyzed the 3400
sq. deg. of the SCUSS footprint that overlaps with SDSS. As can be
seen from this figure, the overlap region is not contiguous due to the
incomplete SDSS coverage for ${\rm RA} > 30^\circ$. The resulting depth
varies somewhat across the footprint, but for most fields the limiting
magnitude lies in the range 23 to 24 mag. Completeness for a given
field is related to the turn-over in this luminosity function, which is
typically 0.5 mag brighter than the limiting magnitude. We have also
measured the location of this turn-over for each field and find that
only 2 per cent have a value brighter than 21.5 mag. For a BHB star
this magnitude corresponds to a distance of around 100 kpc, so we
are confident that our BHB samples should be reasonably complete to
100 kpc and are able to probe significantly further than this
(albeit at lower completeness).

\section{BHB selection}\label{selection}
The class of A-type stars, which includes BHBs, can be selected using
a color-color box in the space of (${u-g}$)$_0$ vs (${g-r}$)$_0$.
In the following work all magnitudes are corrected for extinction
(labeled with a subscript `$_0$') using the maps of
\citet{1998ApJ...500..525S} and adopting the reddening conversions
from \citet{2011ApJ...737..103S}. For SCUSS objects 
with $16<u_0<22.5$ and $ugri$ error less than 0.2, the A-type stars are 
selected following \citep{2000ApJ...540..825Y,2004AJ....127..899S},
\beq
0.9<(u-g)_0<1.4,\\ 
-0.3<(g-r)_0<0.0.
\label{color1}
\eeq
The color-color criterion is efficient at removing contamination of
non-A-type main-sequence stars, white dwarfs and most of the quasars.
To further eliminate quasars from the A-type sample we use the
following color cut \citep[similar to][]{2012MNRAS.425.2840D},
\beq
(g-r)_0 > 0.6164 (g-i)_0 - 0.016.
\label{eq:color2}
\eeq
This cut removes many spurious points at the left and right edges of
the (${u-g}$)$_0$ vs (${g-r}$)$_0$ diagram, making the BHB claw look
more distinct. By cross-matching with SDSS spectroscopy we found that
many of the points removed by the color cuts in Equation
(\ref{eq:color2}) are indeed quasars. In order to check how many
potential quasars are left in the resulting sample, we apply the
XDQSO\footnote{http://www.sdss3.org/svn/repo/xdqso/tags/v0$_{-}$6/doc/build/html/index.html}
technique \citep{2011ApJ...729..141B,2011AnApSB}, which has been
designed to use for the SDSS quasar targeting selection. We found
that the number of quasars (with probability greater than 0.8) left in
our A-type sample is negligible, with a contamination fraction of less
than 2\%. \citet{2012MNRAS.425.2840D} also demonstrated the efficacy
of their $gri$ cut, finding that their spectroscopic sample of 19
faint ($20<g<21.5$) BHBs candidates contained no quasar contamination
when this cut was applied. It is possible that there could be some
variable star contamination (e.g. RR Lyraes) in the current sample,
although it has been estimated that the contamination should only be
around 5\% and hence this will have little bearing on our results
\citep{2011MNRAS.416.2903D}. Also note that since $u_0<$ 22.5, which is
equivalent to a depth of $r\approx$ 21.5, the SDSS star-galaxy separation
should be robust \citep[][see also the SDSS website\footnote{\url{https://www.sdss3.org/dr9/imaging/other_info.php}}]{2001ASPC..238..269L}.

\begin{figure*}
\begin{center}
         \includegraphics[angle=360,width=0.9\textwidth]{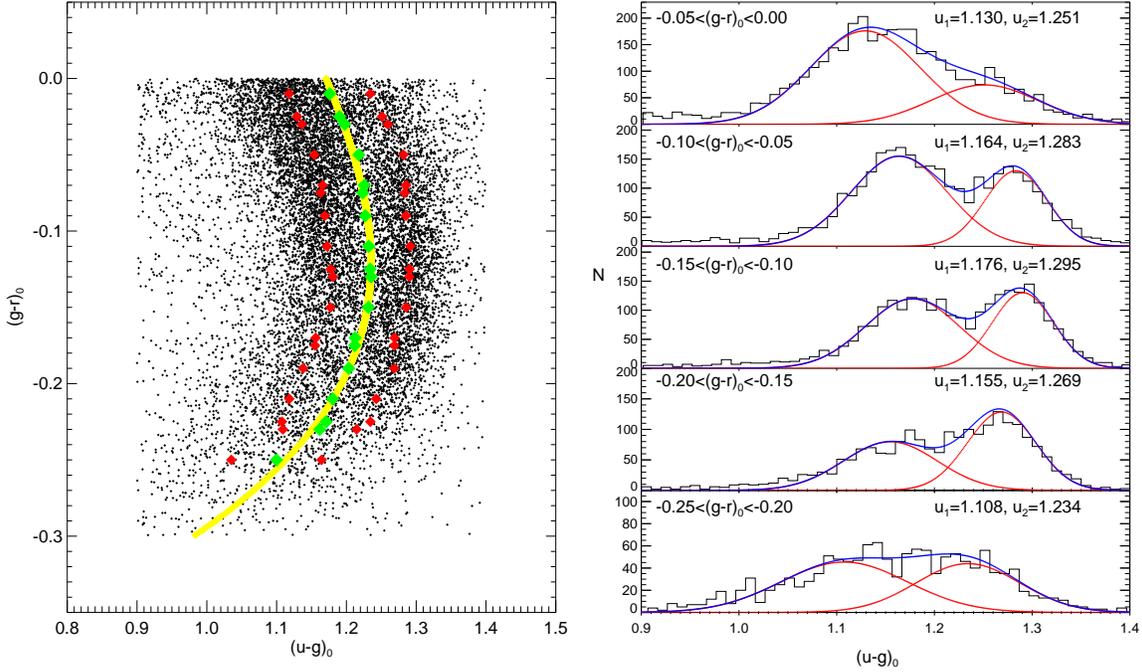}
         \caption{
Color--color selection for BHB stars with $16<u_0<21$ and
$\sigma(u,g,r)<0.1$ mag. Left panel: $(u-g)_0$ vs $(g-r)_0$ diagram
for A--type stars. Red diamonds are the centers of BHB/BS Gaussians
and green diamonds are the mid-point between these values. The yellow 
curve is our fit to these green diamonds. Stars on the right side of 
this yellow curve are considered to be BHB stars.Right panel: An example 
of the two--Gaussian fit for each slice in $(g-r)_0$. Black histograms 
are the $(u-g)_0$ distribution for all A--type stars in each $(g-r)_0$ 
slice, red curves are Gaussian fits to the BHB/BS populations and the blue 
curves are the sum of these two Gaussians.
\label{BHB}}
\end{center}
\end{figure*}

To select BHB stars from A-type sample, we need to separate
them from Blue Stragglers (BS). BS stars also lie in the A-type regime and
so are included within our (${u-g}$)$_0$ vs (${g-r}$)$_0$ selection
(Equation \ref{color1}). To discriminate between BHB/BS stars, the
ideal approach is to analyze the stars' spectra, because the two groups
have different surface gravities and spectral line profiles
\citep{1994AJ....108.1722K,2002MNRAS.337...87C,2004AJ....127..899S,2008ApJ...684.1143X,2011MNRAS.416.2903D}.
Using spectra from SDSS, \citet{2011MNRAS.416.2903D} separated
BHB and BS stars and pinpointed the loci of the two populations in the
(${u-g}$)$_0$ vs (${g-r}$)$_0$ color plane. From their color-color
diagram, the two groups have a distinct division along the $(u-g)_0$
direction (see Figure 2 of \citealt{2011MNRAS.416.2903D}). This
phenomenon is a result of the different surface gravities of BHB 
and BS stars; since the $u$-band filter is located blue-ward of the
Balmer discontinuity, the $(u-g)_0$ color characterizes the strength of
the Balmer jump, a quantity which is sensitive to the surface gravity.
\citet{2011MNRAS.416.2903D} found that the $(u-g)_0$ distributions of
BHB/BS stars can be fit using one Gaussian for each population, with
the center of each Gaussian varying with $(g-r)_0$. By dividing the
color-color plane into several slices in $(g-r)_0$, and modeling the
BHB/BS distributions of each slice with a two-Gaussian function, one
can easily obtain a boundary line for BHB/BS stars.

We classify BHB/BS stars using the above method. However, we cannot
directly apply the boundary line of \citet{2011MNRAS.416.2903D} as the
$u$ filters of SCUSS and SDSS have slightly different profiles
\citep{Zhou2015,Zou2015} and hence objects will have different $(u-g)_0$ 
colors. In order to get an optimal classification for BHB/BS stars we 
determine our boundary line using only bright stars with accurate photometry, 
namely $16<u_0<21$ and $\sigma(u,g,r)<0.1$ mag. We divide the data
along the $(g-r)_0$ direction, with widths of 0.02 mag and 0.05 mag.
This gives us 18 useful slices which we then fit using two Gaussians,
\beq
N = N_1~\rm{exp} \left( \frac{-\left[(u-g)_0-{\mu}_1\right]^2}{2{\sigma_1}^2} \right)
+ N_2 ~\rm{exp} \left( \frac{-\left[(u-g)_0-{\mu}_2\right]^2}{2{\sigma_2}^2} \right).
\eeq
The fitting procedure is illustrated in the right panel of Figure
\ref{BHB}, where we show bin widths of 0.05 mag. The two separate
populations can clearly be discerned, with BHB stars lying to the 
red (i.e. right) side of the distribution. By comparing this figure 
to the similar one from \citet{2011MNRAS.416.2903D}, it is evident 
that the SCUSS $u$-filter is much more gravity-sensitive, as the BHB/BS 
populations exhibit less overlap and, in a number of $(g-r)_0$ slices, 
there are clear minima in the $(u-g)_0$ distributions. The 
division is not perfect, meaning that BHB samples will be incomplete and 
contaminated by BSs (see Section \ref{pisces}), but it is clear that the 
SCUSS $u$-filter performs better in this regard compared to the SDSS 
$u$-filter.

The fits for all 18 slices are shown in the left panel of Figure
\ref{BHB}, where the red points denote the centers of the Gaussians
($\mu_1$ and $\mu_2$), corresponding to the centers of the BS and BHB
populations, respectively. We adopt the median of these two values
(i.e. $\mu=(\mu_1+\mu_2)/2$; green points in Figure \ref{BHB}) as the
boundary between the two populations. We have chosen $\mu$ to
represent the boundary line because this quantity is magnitude
independent. For a given $(g-r)_0$ slice the ratio of BHB to BS stars ($N_1/N_2$) 
will vary as a function of magnitude, which means that the overall shape 
of the $(u-g)_0$ distribution will alter; however, even though the shape 
will change, it can be seen that $\mu$ is independent of magnitude. In 
order to calculate our expression for the boundary line, we fit the
values of $\mu$ using the following polynomial function, as shown by the 
yellow curve in Figure \ref{BHB}:
\beq
\label{boundary}
{(u-g)}_{\rm b}=\\
1.171-0.888(g-r)_0-1.531{(g-r)_0}^2+11.791{(g-r)_0}^3
\eeq
where stars on the right side of this curve are taken to be BHB stars.

We now apply this classification to the whole magnitude range
($16 < u_0 < 22.5$), taking all stars with errors in
$ugri$ of less than 0.2 mag. As mentioned above, even
though Equation (\ref{boundary}) was derived for stars with $16<u_0<21$,
we are able to apply it now to the full magnitude range because $\mu$
should be independent of magnitude. BHB distances are 
calculated using Equation (7) of \citet{2011MNRAS.416.2903D}, which presents 
a relation between absolute-magnitude and $(g-r)_o$ color with $M_g \approx 0.5$ mag.

\section{BHB stars around Pisces}
\label{pisces}

Given this large and relatively clean sample of BHBs, we now proceed
to investigate the Pisces overdensity. 

We take BHBs according to the boundary defined in Equation
(\ref{boundary}), but in order to further reduce contamination we
reject all stars within 0.02 mag of the boundary and also those which
are more than 0.15 mag away, i.e. we only retain stars with
$0.02 < \Delta(u-g)_0 = (u-g)_0 - (u-g)_{\rm b} < 0.15$ mag.
The latter cut does not remove many stars, but the former is important
for reducing the number of BS contaminants in the sample.
Note that although the error cut is 0.2 mag in each
band, our final errors are considerably smaller. For example, the
median error on $(u-g)$ for our entire sample is 0.032 mag. Even at
faint magnitudes the errors are reasonable, with BHBs at 100 kpc
having median error of 0.1 mag.

Since our photometric precision is good and the SCUSS $u$-filter is
more sensitive to gravity than the SDSS filter, the level of BS
contamination is low. However, as can be seen from the right panel of
Figure \ref{BHB}, where it is clear that the BHB and BS Gaussians
overlap, this is not negligible and must be estimated.
For the typical distances we wish to investigate (i.e. heliocentric
distances of 35 to 90 kpc, corresponding to $19.5<u_0<21.5$ mag), we
find that the fraction of BSs contaminating our sample is between 8\%
and 18\%, depending on $(g-r)_0$. As there are no known structures in the
foreground of the Pisces region, these BSs should be uniformly
distributed and therefore this level of contamination is unlikely to be 
problematic. Note also that our sample will not be complete, as a
number of BHBs lie to the left of the boundary line. We estimate that
our BHB sample is around 80\% complete.

Now we have our sample of BHBs, we use a standard K-nearest-neighbor
algorithm to locate the density peaks. This technique simply
calculates the density at a point by averaging over a volume which
includes the K nearest neighbors, where the choice of K depends on the
problem at hand (in effect this is a smoothing length). For our work
we have chosen a value of K$=$10, although our results are similar if
one takes K$=$8 or 12.
When calculating the density we have subtracted
a smooth background model, using a flattened ($q=0.7$) double
power-law profile fit to the RR Lyrae data of
\citet{2009MNRAS.398.1757W}; the fit parameters are given in
\citet[][2nd row of Table 4]{Faccioli2014}.
Since this profile corresponds to RR
Lyrae and not BHBs, we have increased the normalization by a factor of
2, which we chose in order to match our observed distribution of
BHBs. Given this smooth background model, we can calculate the significance 
of any volume in our 3D space, calculating the difference between
the observed and predicted number of BHBs.

\begin{figure*}
\includegraphics[width=18cm]{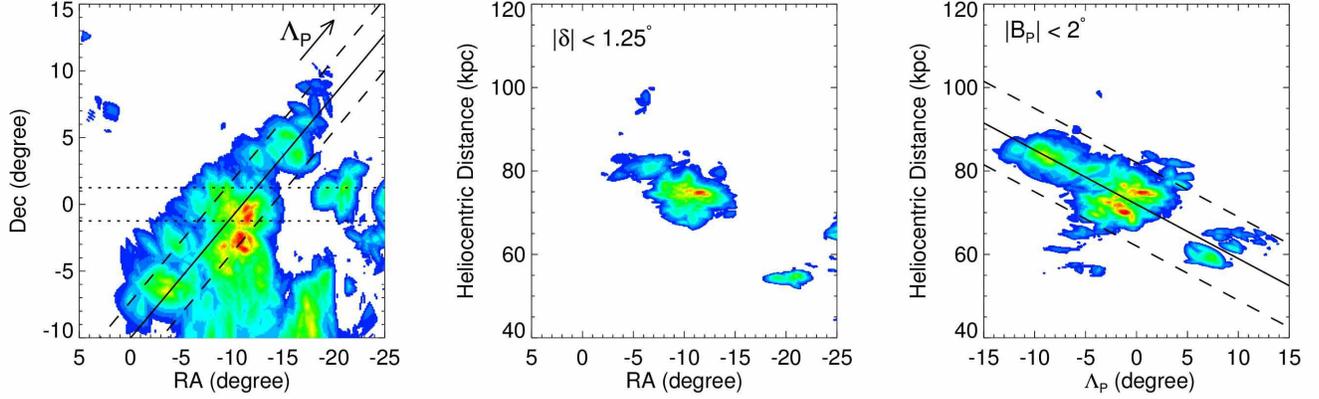}
\caption{The density of BHBs around the Pisces overdensity,
calculated using an 10-nearest-neighbor method. The original detection
was in the narrow Stripe 82 region (dotted lines in the left panel) at
$-5^\circ<{\rm RA}<-15^\circ$ and the corresponding distances to
our BHBs overdensities in this region are shown in the middle
panel. The left panel shows the BHB density for areas surrounding the
original detection. The overdensity shows an extension
along a stream, denoted by the black solid line (see Appendix for the
coordinate transformation). The stream plane is defined by two angles
($\Lambda_{\rm P}$ and ${\rm B_P}$), where $\Lambda_{\rm P}$ is
oriented along the stream and increases with decreasing RA, and
${\rm B}_{\rm P}$ is oriented across the stream and increases with
increasing Dec. The dashed lines lie 2 deg either side of this path
and the distances of BHBs within this region are shown in the right panel. 
The fact that the distances lie along a well-defined distance gradient confirm 
our interpretation that Pisces is part of a stream.}
\label{fig:stream}
\end{figure*}

We present our map of the BHB density distribution in the left panel
of Figure \ref{fig:stream}. Since the Pisces overdensity was discovered 
in the SDSS equatorial Stripe 82 data, we first investigate that region
(i.e. $\left|~{\rm Dec}~\right|<1.25^\circ$). Indeed there is a strong 
overdensity located around $-5^\circ<{\rm RA}<-15^\circ$, which is precisely 
where it was previously identified. The other features of note in this
region are the additional structures at $-25^\circ <{\rm RA}<-20^\circ$, but 
these are most-likely the edge of the Hercules-Aquila 
Cloud \citep{2007ApJ...657L..89B,Simion2014}, which was also detected
in Stripe 82 \citep[e.g.][]{2009MNRAS.398.1757W}.

When we compare the distance to our Pisces detection (middle panel of
Figure \ref{fig:stream}; again restricting ourselves to just the
Stripe 82 region), we find a weak secondary detection behind the main
clump at around 95--100 kpc. This more-distant feature has only 5
members and hence may be a spurious detection, but the significance is
noticeable (at around 2$\sigma$) because the model predicts very few
stars this far out in the halo. The detection is unlikely to be caused by 
extra-galactic contamination;  XDQSO classifies all 5 members as stars and 
so the overdensity is unlikely to be caused by background quasar contamination
and, since $r\approx20.5$ mag for these stars, the SDSS star-galaxy 
classification should be robust \citep{2001ASPC..238..269L}. This detection 
cannot be a `shadow' of mis-classified BSs from the main structure, since the 
offset in distance modulus is around 0.6 mag, not the 2 mag that one would 
expect if these were misclassified BSs. Also, with the exception of one star, 
all are far from the BS/BHB boundary and so this reinforces our belief that 
these are not BS contaminants.

If one compares our distance distribution to the RR Lyrae detection
from \citet[][Figure 16]{2009MNRAS.398.1757W}, then one also sees a
hint of bimodality and a similar spread in distances (around 40 kpc).
It should be noted that the peak of our BHB detection is slightly
offset from the RR Lyrae detection, with distances of 75 and 80 kpc,
respectively. Since there is more uncertainty in the BHB distance
calibration (as opposed to the RR Lyrae distances), it may be that
this is out by 0.1 to 0.2 mag. This could be due to deficiencies
in our adopted absolute-magnitude relation
\citep[Equation 7][]{2011MNRAS.416.2903D}, such as a metallicity bias
\citep{Fermani2013}, or differences in the adopted extinction values.

The next step is to expand our search beyond the Stripe 82
region. As can be seen the left panel of Figure \ref{fig:stream},
the wide coverage of the SCUSS data, combined with the clean
separation of BHBs from BSs, allows us to tentatively detect an
extension of the overdensity to both lower and higher
declinations. These manifest themselves as separate clumps, which seem
to be connected to the peak overdensity lying at RA $= -10^\circ$ 
and Dec $= -3^\circ$. The significance of these two regions is 3.5 and
3.0$\sigma$, respectively, for the lower- and higher-declination regions.
There are some additional clumps at $-25^\circ <{\rm RA} < -20^\circ$, 
but again these are most-likely the edge of the Hercules-Aquila Cloud.

It seems that the Pisces overdensity is extended like a stream,
passing from the bottom-left to top-right of the figure. In order to
confirm this we now analyze the distances of the BHBs in this
feature. We define a path for the stream (denoted $\rm \Lambda_P$ and
${\rm B_P}$; see Appendix) and take BHBs within 2 deg. The density map 
of these is shown in the right panel of Figure \ref{fig:stream}. Here we 
can see that the detections are aligned beautifully along a distance gradient, 
confirming our interpretation of a stream. Again the curious cloud of distant 
BHBs is noticeable beyond 100 kpc, but is clearly unrelated to the stream. The
distance gradient along the stream is around 1.3 kpc per deg,
\begin{equation}
{\rm D_P = 1.3\times \Lambda_P} + 72,
\end{equation}
where $D_P$ is the helio-centric distance of the stream in kpc.

We calculate the significance by taking a volume encompassing our 
Pisces detection $(\rm -12 < \Lambda_P < 10~deg, \left| B_P \right| <
2~deg,~\left| D-D_P \right| < 10~kpc)$ and calculating the number of
BHB stars predicted from the smooth model, then compare this to the
observed number. The model predicts that there should be 8--9 stars in 
this volume, while we actually find 27 stars. This implies that our
detection is 6.2$\sigma$. The significance varies depending on
how one defines the background model, for example the flattening or
the normalization, but the variation is small and so we conclude that
the significance is around 5--7$\sigma$. The significance of the new
detections that extend away from the central region, i.e. at 
$({\rm -12 < \Lambda_P < -5 ~deg})$ and $({\rm 3 < \Lambda_P < 10~deg})$, 
is 3.5 and 3.3$\sigma$, respectively.

Although we find that 27 stars belong to our detection, this is only an
approximate number. As discussed above, some of these may be BS
contaminants, while some bona fide BHBs may lie beyond our BHB/BS
boundary line. An additional four objects were rejected by our
quasar cut (Equation \ref{eq:color2}), of which two are likely to be
stars according to XDQSO. On the other hand, of the 27 objects that
passed our quasar cut, one was classified as a quasar by XDQSO.

There are a couple of notes of caution which should be
addressed. Firstly, there appears to be a gap in the stream at
$\Lambda_P \sim 4$ deg and $\rm D_P \sim 65$ kpc. We have checked
whether this could be due to bad fields or patchy extinction, but
neither are found at this location. However, the density of BHBs is
low and so this gap could just be due to statistical fluctuations. The
other cautionary point is the clump of material around $-14^\circ <
{\rm RA}< -8^\circ$ and $-10^\circ < {\rm Dec}< -5^\circ$.
Although this is statistically significant (at around 3$\sigma$), 
the distance distribution is very broad and there is no obvious
clumping at the distance of Pisces.

We have chosen to focus on the region around the Pisces overdensity,
even though the SCUSS footprint is much larger. However, if we
extend the area to the entire footprint, we do not detect
any significant new structures.

\section{Giant branch exploration}\label{giant_branch}

\begin{figure*}
\includegraphics[width=18cm]{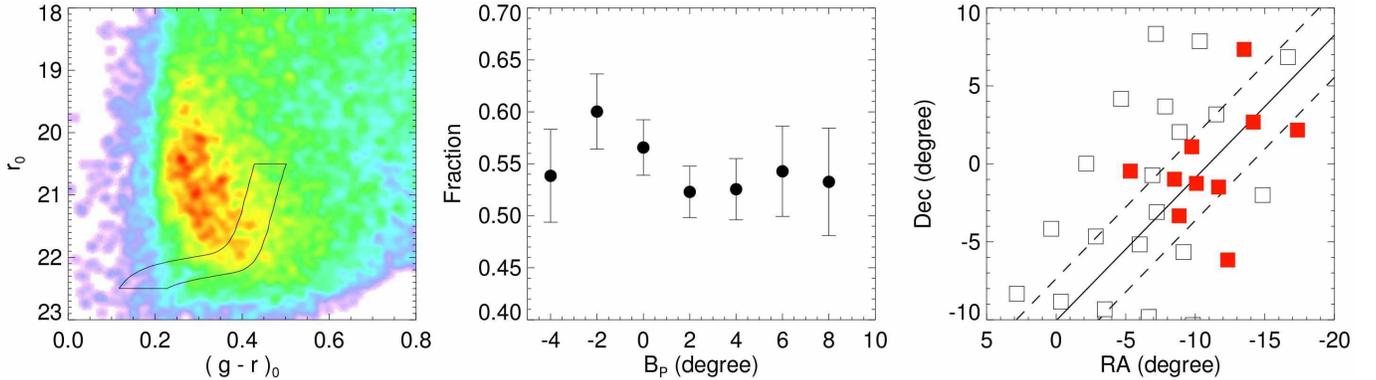}
\caption{Our search for the giant branch of the Pisces
overdensity, using data from CFHT. The left panel shows a Hess diagram
for all of our fields, with the box corresponding to an isochrone for
an 8 Gyr population with $[Fe/H] = -1.5$ dex, located at 75 kpc. To
give the box a finite size we have shifted it by $\pm 5$ kpc and by
$\pm 0.03$ mag in $(g-r)_0$. The middle panel shows the number of
stars within the isochrone box, normalized by the total number in the
same magnitude range ($22.5<r_0<20.5$, $0 < (g-r)_0 < 0.8$) but excluding
stars inside the isochrone box and those within 0.1 mag in $(g-r)_0$
of the box boundary. This is shown as a function of cross-stream angle
$\rm B_P$ and errors are assumed to be Poissonian. Finally, in the right
panel we show the location of our CFHT fields. Filled squares are
those which have fractions greater than the background level by at
least 0.5$\sigma$.}
\label{fig:giant_branch}
\end{figure*}

In an attempt to compliment this analysis, we have also analyzed data
obtained from the 3.6m Canada-France-Hawaii Telescope (CFHT) using the
1-square-degree field-of-view MegaPrime/MegaCam camera. We chose 8 min
exposures in $r$ (seeing of $\sim$0.5$\arcsec$) and 12 min in
$g$ (seeing of $\sim$0.9$\arcsec$), allowing us to reach around 24
mag in both bands. Since we could not contiguously map the full
region around Pisces, we limited ourselves to 30 pointings (i.e. 30
sq. deg.) spread over a wide area. The location of these fields are
shown in the right panel of Figure \ref{fig:giant_branch}. Although
the data reach to 24 mag, the star-galaxy separation (which 
is dependent on the seeing and signal-to-noise ratio) becomes problematic 
below around $r\approx22$ mag. In order to alleviate this problem we cross-match 
with SCUSS $u$ band and apply the following mask to isolate the stellar locus
\citep[see, for example][]{Strateva2001}
\beq
\left|~
(u-r)_0 - 0.37 (g-r)_0
~\right|
< 0.75,
\eeq
and also apply $(u-r)_0 > 0.75$ to remove quasars.

The resulting Hess diagram for our data is shown in the left panel of
Figure \ref{fig:giant_branch}. The main-sequence turn-off for the
stellar halo is clearly visible, but there is no obvious detection of
any sequence belonging to the Pisces overdensity. To guide the eye, we
have included in this figure a box corresponding to the isochrone for
an 8 Gyr population with $[Fe/H] = -1.5$ dex, located at 75 kpc. To
give the box a finite size we have shifted it by $\pm 5$ kpc and by
$\pm 0.03$ mag in $(g-r)_0$. Due to the limitations of the SCUSS data, 
completeness begins to drop around $r\approx22.5$ mag and we are
therefore unable to detect the main-sequence turn-off. If the giant
branch is present, it is not immediately obvious although it may be
hidden behind the main-sequence of the stellar halo, which is
considerably more dense.

We investigate further by calculating the number of stars within the
isochrone box shown in Figure \ref{fig:giant_branch}. In order to
account for variations in the background density of stars across our
fields, we need to normalize this number. We do this by dividing the
number of stars inside the isochrone box $\rm N_{in}$ by 
$\rm N_{out}$, where $\rm N_{out}$ is the total number of stars in the
same magnitude range ($22.5<r_0<20.5$, $0 < (g-r)_0 < 0.8$) {\it
excluding} stars inside the isochrone box and those within 0.1 mag in
$(g-r)_0$ of the box boundary. We calculate this fraction 
$\rm N_{in}/N_{out}$ as a function of $\rm B_P$, the cross-stream angle 
introduced in the previous section and plot this in the middle panel of 
Figure \ref{fig:giant_branch}. At most locations this fraction is relatively
constant at around 0.425, but for two bins close to the centre of the
stream the value is notably above this (at a significance of 1- to
2-$\sigma$, where errors are assumed to be Poissonian).

Finally we split our data into the individual CFHT fields. We first
estimate the background by averaging all $\rm B_P$ bins except the two
at $\rm B_P=-2^\circ$ and 0$^\circ$. We then calculate this fraction for 
each of our fields and, in the right panel of Figure \ref{fig:giant_branch}
show with filled boxes the fields for which the fraction is at least
0.5$\sigma$ above the background value. From this figure one can see
that there is a tendency for fields with larger fractions to lie
within 2 deg. of the proposed stream plane (shown by the dashed
lines). However, this is inconclusive, with most fields having excess
at only 1- to 2-$\sigma$.

In summary, our search for the Pisces giant branch shows tentative
evidence for alignment with the stream plane in the previous section,
but the results are inconclusive and velocities would be required to
clarify the situation.

\section{Discussion}\label{conclusion}

In this paper we have provided an extended view of the Pisces
overdensity, utilizing deep $u$ band data from the SCUSS survey.   
With the SCUSS $u$ band data, which goes around 1 to 1.5 mag deeper than
single-epoch SDSS, we have used BHB stars as tracers and analyzed the
distribution of overdensities for $\sim$1000 sq. deg. around the
original Pisces detection. We have found that Pisces appears to be
part of a stream, with a clear distance gradient (Figure
\ref{fig:stream}). This stream is around 5 deg. wide and we appear to
trace it for 25 deg. in length. Given the considerable distance
to this stream (60 to 80 kpc) this makes it the largest
structure (by volume) in the outer halo after the tidal tails of the
Sagittarius dwarf galaxy. We find that there are 27 BHBs
associated to our extended detection of the overdensity, although we
cannot report this number with any high degree of accuracy due to
three effects: firstly, some of these BHBs will not belong to the
overdensity and are either smooth halo BHBs or foreground BSs;
secondly some BHBs will be lost due to incompleteness, especially
those which have smaller (bluer) $(u-g)_0$ and are 
hence indistinguishable from BSs; and thirdly, there may be member stars 
outside the $\Lambda_{\rm P}$ range we have detected. 
We can roughly estimate the influence of the first two effects,
using the contamination and completeness fractions calculated in
Section \ref{pisces}. After accounting for these, we predict that
total number of BHBs belonging to our Pisecs detection is around 21
stars. With this rough number in hand, we can speculate as to the
progenitor of the stream. Assuming we have detected the entire stream
and that the progenitor has fully disrupted, then this number of BHBs
suggests that the progenitor was likely somewhere between a smaller
classical dwarf spheroidal such as Draco or Sextans
\citep[e.g.][]{Aparicio2001,Lee2003}, and a larger ultra-faint dwarf
like Canes Venatici I \citep[e.g.][]{Okamoto2012}.

Compared to the other Pisces detections, which are restricted to the
Stripe 82 region \citep[][]{2007AJ....134.2236S,2009ApJ...705L.158K,2009MNRAS.398.1757W,2010ApJ...717..133S}, our detection of a tangential extension is new. Before, 
the only tentative evidence of an extension was reported by \citet{2010ApJ...722..750S}. 
Using M-giant tracers and a group finding method, they report that the 
Pisces overdensity extends from $-20^{\circ}<$RA$<25^{\circ}$,
$1.25^{\circ}<$Dec$<25^{\circ}$, with a distance of 103$\pm$51
kpc (see Figure 9 of their paper). Compared to our detection,  
there is a similarity in the region of $-20^{\circ}<$RA$<-10^{\circ}$,
$1.25^{\circ}<$Dec$<11^{\circ}$, but their detection at RA$>-10^\circ$
is totally different from ours; here we find the overdensity extending
to negative declinations, but the \citeauthor{2010ApJ...722..750S}
detection extends to larger positive declinations. Furthermore, this
detection does not overlap with the Stripe 82 detection. A more-recent
M-giant map of that region is presented \citep{Deason2014}. Although
not remarked on by the authors themselves, an inspection of the left
panel of their Figure 5 shows an overdensity at the same location as
the Pisces detection, seemingly oriented in the same direction as
our BHB detection.

The only discrepancy between our results and the existing Stripe 82
detections is the offset in distance, with ours being around 10 \%
smaller than those estimated from RR Lyrae. We are unable to determine
the cause of this offset. Like \citet{2009MNRAS.398.1757W} we also
detect a hint that the Stripe 82 detection may be split into a near
and far component, but further work needs to be carried out to confirm
this. In particular is it unclear whether this could explain the 
multiple kinematic groups in the Stripe 82 detection, as claimed by 
\citet{2009ApJ...705L.158K} and \citet{2010ApJ...717..133S}. It
appears that the distances to both features are similar \citep[see, for
example, Figure 1 of][]{2010ApJ...717..133S}, but with such small
numbers of stars it is hard to make firm conclusions. It is also clear 
that our detection of the stream is rather `lumpy', which may reflect the 
lumpiness of the stream itself, or possibly due to the sparseness of the 
tracer population.

Since we have detected a stream, it is natural to ask whether the core
of the progenitor remains intact. There is a dwarf galaxy in the
vicinity of the stream, the recently detected Pisces II dwarf
(RA=344.6$^\circ$, Dec=5.9$^\circ$) \citep{2010ApJ...712L.103B}. If we
transform its sky coordinates to our stream coordinates, we find that
it has $\Lambda_{\rm P}=7.9^\circ$ and $\rm{B_P}=0.7^\circ$. However,
despite it's close proximity (in projection) to the stream plane, its
distance of 180 kpc precludes it from being the progenitor, at least
not if it is part of the same wrap of the stream. Given our detected
distance gradient, it is now possible to model the orbit of the stream
and embark on a large scale hunt for the progenitor.

\acknowledgments 
The authors would like to thank Alis Deason, Jo Bovy and the anonymous
referee for contributions to the study.

This work is supported by the following sources: the National Natural
Science Foundation of China (11303043, 11173002, 11333003, 11203031,
11203034, 11303038, 11373003, 11373033, 11373035, 11433005); the
National Basic Research Program of China 973 Program (2013CB834902,
2014CB845700, 2014CB845702, 2014CB845704); the Chinese Academy of
Sciences (CAS) Strategic Priority Research Program "The Emergence of
Cosmological Structures" (XDB09000000); and the Gaia Research for
European Astronomy Training (GREAT-ITN) Marie Curie network, funded
through the European Union Seventh Framework Programme (FP7/2007-2013)
under grant agreement no 264895. M.C.S. acknowledges financial support
from the CAS One Hundred Talent Fund.

SCUSS is funded by the CAS Main Direction Program of Knowledge
Innovation (KJCX2-EW-T06). It is also an international cooperative
project between National Astronomical Observatories (CAS) and Steward
Observatory (University of Arizona, USA). Technical support and
observational assistances of the Bok telescope are provided by Steward
Observatory. The project is managed by the National Astronomical
Observatory of China and Shanghai Astronomical Observatory.

This paper uses data obtained through the Telescope Access Program
(TAP), which has been funded by the Strategic Priority Research
Program "The Emergence of Cosmological Structures" (Grant
No. XDB09000000), National Astronomical Observatories, Chinese Academy
of Sciences, and the Special Fund for Astronomy from the Ministry of
Finance.

This paper uses data from SDSS-III. Funding for this project has been
provided by the Alfred P. Sloan Foundation, the Participating
Institutions, the National Science Foundation, and the U.S. Department
of Energy Office of Science. The SDSS-III web site is
\url{http://www.sdss3.org/}. SDSS-III is managed by the
Astrophysical Research Consortium for the Participating Institutions
of the SDSS-III Collaboration including the University of Arizona, the
Brazilian Participation Group, Brookhaven National Laboratory,
Carnegie Mellon University, University of Florida, the French
Participation Group, the German Participation Group, Harvard
University, the Instituto de Astrofisica de Canarias, the Michigan
State/Notre Dame/JINA Participation Group, Johns Hopkins University,
Lawrence Berkeley National Laboratory, Max Planck Institute for
Astrophysics, Max Planck Institute for Extraterrestrial Physics, New
Mexico State University, New York University, Ohio State University,
Pennsylvania State University, University of Portsmouth, Princeton
University, the Spanish Participation Group, University of Tokyo,
University of Utah, Vanderbilt University, University of Virginia,
University of Washington, and Yale University.

\appendix
\section{The coordinate transformation}
\label{app:plane}
Here, we provide the equations for converting from Equatorial ($\alpha$,
$\delta$) to the Pisces Overdensity coordinate system ($\Lambda_{\rm
P}$, B$_{\rm P}$). The orbital pole of the Pisces overdensity is set to
($\alpha_{P}$, $\delta_{P}$), and its center is set to ($\alpha_c$,
$\delta_c$). Details of the derivation are shown below.

Spherical coordinates can be converted to a right-handed
Cartesian coordinate system using,
\begin{displaymath}
\left (\begin{array}{c}
\rm X\\
\rm Y\\
\rm Z\\
\end{array}
\right)=\left(\begin{array}{c}
\cos\alpha \cos\delta \\
\sin\alpha \cos\delta \\
\sin\delta \\
\end{array}
\right)~~.
\end{displaymath}
The orbital pole of the Pisces system ($\alpha_{P}$,
$\delta_{P}$) can be rotated into the new Cartesian coordinate system
as follows,
\begin{displaymath}
\left (\begin{array}{c}
       \rm	X^{'}\\
       \rm	Y^{'}\\
       \rm	Z^{'}\\
\end{array}
\right)=
\left (\begin{array}{c}
R_x~\rm X\\
R_y~\rm Y\\
R_z~\rm Z\\
\end{array}
\right)~~,
\end{displaymath}
where
\[R_{x}=\frac{A_1\times A_2}{\mid A_1 \times A_2\mid},~~A_1=[\cos\alpha_{P} \cos\delta_{P},~ \sin\alpha_{P} \cos\delta_{P},~\sin\delta_{P}],~A_2=[0,0,1]~,
\]
\[
	R_{y}=\frac{R_{x}\times A_1}{\mid R_{x} \times A_1\mid}~~,
\]
\[
R_{z}=[	\cos\alpha_{P} \cos\delta_{P},~ \sin\alpha_{P} \cos\delta_{P},~\sin\delta_{P}].
\]
With $\rm X^{'}$,$\rm Y^{'}$ and $\rm Z^{'}$, 
\[\Lambda_{\rm{P}}=\rm{atan2}(Y^{'},X^{'})
\]
and
\[\rm{B}_P=\arcsin(Z^{'}) ~~~~,
\]
where $\tan(\rm{atan2}((Y^{'},X^{'}))=(Y^{'}/X^{'})$.

If the center of the Pisces overdensity is set to ($\alpha_c$,
$\delta_c$), then we need to put this in phase with the original
coordinate system,
\[\Lambda_{\rm{P}}=\Lambda_{\rm{P}}-\Lambda_{\rm{PC}}
\]
\[\rm{B}_P=\rm{B}_P-\rm{B}_{PC}
\]
where $\Lambda_{\rm{PC}}$ and $\rm{B}_{PC}$ is for ($\alpha$, $\delta$)= ($\alpha_c$, $\delta_c$).

Our orbital plane uses ($\alpha_{P}$, $\delta_{P}$) =
(79$^\circ$, 47$^\circ$) ($\alpha_c$, $\delta_c$) = (-10$^\circ$,
0$^\circ$), which results in the following transformation,
\beq
\label{transfermation1}
\begin{split}
\Lambda_{\rm{P}}=\rm{atan2}(-0.13954893\cos \alpha \cos \delta-0.71791667\sin \alpha 
\cos \delta+0.68199837 \sin \delta,~~~~~~~~~~~~~~~~\\
~~~~~~~~~~~~~~0.98162711\cos \alpha \cos \delta -0.19080906\sin \alpha\cos \delta )+0.73139161^\circ~~~~~~~~~~~~~~~~~~~~~~~~~~~~~~~~~~~~~~~\\
\rm{B}_P=\arcsin(0.13013147\cos \alpha\cos \delta +0.66946810\sin \alpha \cos \delta+0.73135370\sin \delta)~~~~~~~~~~~~~~~~~~
\end{split}
\eeq

Note that in this system $\Lambda_{\rm P}$ is oriented along the
stream and increases with decreasing $\alpha$, and ${\rm B}_{\rm P}$ is
oriented across the stream and increases with increasing $\delta$.

The reverse transformation from the Pisces Overdensity coordinate system
($\Lambda_{\rm{P}},\rm{B}_P$) to the Equatorial system ($\alpha,\delta$)
is, 
\begin{equation}
\label{transfermation2}
\begin{split}
	\alpha=\rm{atan2}(-0.19080907\cos \Lambda^\prime_{\rm P}  \cos \rm{B}_P -0.71791661\sin \Lambda^\prime_{\rm P}
\cos \rm{B}_P +0.66946810 \sin \rm{B}_P,
\\
0.98162729\cos \Lambda^\prime_{\rm P} \cos \rm{B}_P -0.13954890\sin
\Lambda^\prime_{\rm P} \cos \rm{B}_P +0.13013147 \sin \rm{B}_P )
~~~~~~~~~~~\\
\delta=\arcsin(0.68199837\sin \Lambda^\prime_{\rm P} \cos \rm{B}_P+0.73135370\sin
\rm{B}_P)
~~~~~~~~~~~~~~~~~~~~~~~~~~~~~~~~~~~~~~\\
\Lambda^\prime_{\rm P} = \Lambda_{\rm P} - 0.73139161^\circ
~~~~~~~~~~~~~~~~~~~~~~~~~~~~~~~~~~~~~~~~~~~~~~~~~~~~~~~~~~~~~~~~~~~~~~~~~~~~~~~~~\\
\end{split}
\eeq

\end{document}